\documentclass{article}

\usepackage[english]{babel}
\usepackage[backend=biber,style=nature, sorting=none]{biblatex}
\usepackage{amssymb}

\usepackage{bbding}

\usepackage[letterpaper,top=2cm,bottom=2cm,left=3cm,right=3cm,marginparwidth=1.75cm]{geometry}

\usepackage{amsmath}
\usepackage{graphicx}

\usepackage[colorlinks=true, allcolors=blue]{hyperref}
\usepackage{subcaption,booktabs}
\usepackage{xcolor,colortbl}
\definecolor{lavender}{rgb}{0.9, 0.9, 0.98}

\usepackage{siunitx}

\providecommand{\keywords}[1]
{
  \small	
  \textbf{\textit{Keywords---}} #1
}

\bibliography{sample.bib}

\title{Deep-learning-driven end-to-end metalens imaging}

\usepackage{authblk}

\author[1,${\dagger}$]{Joonhyuk Seo}
\author[2,${\dagger}$]{Jaegang Jo}
\author[3,${\dagger}$]{Joohoon Kim}
\author[4,${\dagger}$]{Joonho Kang}
\author[1]{Chanik Kang}
\author[3]{Seongwon Moon}
\author[5]{Eunji Lee}
\author[1,2,4]{Jehyeong Hong}
\author[3,5,6,7,8,$*$,\Envelope$^1$]{Junsuk Rho}
\author[1,2,4,$*$,\Envelope$^2$]{Haejun Chung}

\affil[1]{Department of Artificial Intelligence, Hanyang University, Seoul, 04763, Republic of Korea}
\affil[2]{Department of Electronic Engineering, Hanyang University, Seoul, 04763, Republic of Korea}
\affil[3]{Department of Mechanical Engineering, Pohang University of Science and Technology (POSTECH), Pohang, 37673, Republic of Korea}
\affil[4]{Graduate School of Artificial Intelligence Semiconductor, Hanyang University, Seoul, 04763, Hanyang University}
\affil[5]{Department of Chemical Engineering, Pohang University of Science and Technology (POSTECH), Pohang, 37673, Republic of Korea}
\affil[6]{Department of Electrical Engineering, Pohang University of Science and Technology (POSTECH), Pohang, 37673, Republic of Korea}
\affil[7]{POSCO-POSTECH-RIST Convergence Research Center for Flat Optics and Metaphotonics, Pohang, 37673, Republic of Korea}
\affil[8]{National Institute of Nanomaterials Technology (NINT), Pohang, 37673, Republic of Korea}
\affil[${\dagger}$]{These authors contributed equally to this work.}
\affil[$*$]{These authors are corresponding authors.}

\begin{document}
\maketitle

\begin{center}
    $1, 2$ \{yhy258, jaegang0626, joonhokang12, chanik, jhh37, haejun\}@hanyang.ac.kr
    
    $3, 4$ \{kimjuhoon, swmoon, ejlee0315, jsrho\}@postech.ac.kr
    
    \Envelope$^1$ jsrho@postech.ac.kr, +82-54-279-2187
    
    \Envelope$^2$ haejun@hanyang.ac.kr, +82-2-2220-3601
\end{center}

\begin{abstract}
Recent advances in metasurface lenses (metalenses) have shown great potential for opening a new era in compact imaging, photography, light detection and ranging (LiDAR), and virtual reality/augmented reality (VR/AR) applications. However, the fundamental trade-off between broadband focusing efficiency and operating bandwidth limits the performance of broadband metalenses, resulting in chromatic aberration, angular aberration, and a relatively low efficiency. In this study, a deep-learning-based image restoration framework is proposed to overcome these limitations and realize end-to-end metalens imaging, thereby achieving aberration-free full-color imaging for mass-produced metalenses with 10-mm diameter. Neural-network-assisted metalens imaging achieved a high resolution comparable to that of the ground truth image.\\
\keywords{visible metalens, deep learning, image restoration, full color imaging }
\end{abstract}

\section{Introduction}
The  unyielding pursuit of miniaturization and performance enhancement in optical imaging systems has led to the exploration of innovative technologies beyond conventional geometric lens-based systems. While foundational to modern optics, these systems face inherent limitations such as chromatic ~\cite{wang2018broadband,chen2018broadband} and spherical aberrations~\cite{yang2023wide, fan2020ultrawide}, shadowing effects~\cite{fan2020ultrawide, yang2023wide}, bulkiness~\cite{khorasaninejad2016metalenses, shrestha2018broadband}, and high manufacturing costs~\cite{kim2023one, kim2022metasurface}. The quest to transcend these barriers has catalyzed the advent of metalenses, a groundbreaking development poised to redefine the landscape of optical engineering.

Metalenses, characterized by their ultrathin films with meticulously arranged subwavelength structures called meta-atoms interspersed throughout, emerged as a revolutionary alternative to overcome the drawbacks of conventional lenses. In a recent study, deep-ultraviolet immersion lithography was combined with wafer-scale nano-imprint lithography to mass-produce low-cost and high-throughput large-aperture metalenses, contributing to their commercialization~\cite{kim2023scalable}. This novel class of lenses also promises to rectify the aforementioned issues existent in conventional optics and opens a new era of compact, efficient imaging systems~\cite{zhou2020flat,wang2018broadband}. Central to the appeal of metalenses is their ability to serve as optimal substitutes for traditional optical elements and thereby revolutionize a broad spectrum of applications. This encompasses not only the enhancement of capabilities of optical sensors~\cite{shalaginov2020single}, smartphone cameras~\cite{martins2020metalenses, wang2022visible}, and unmanned aerial vehicle optics~\cite{chen2022meta, kim2021dielectric} but also the transformation of user experiences facilitated by augmented and virtual reality devices~\cite{li2021meta, li2022inverse}. The potential of unparalleled diffraction-limited focusing within an ultra-light and ultra-compact form factor, even in high-NA regimes~\cite{liang2018ultrahigh}, a feat unattainable by traditional components, is the key attribute contributing to these advancements.

Despite these strides, the pursuit of broadband metalenses uncovers a multifaceted trade-off among focusing efficiency, lens diameter, and spectral bandwidth~\cite{miller2021fundamental, kamali2018review}, with the last significantly affected by chromatic aberration~\cite{engelberg2017optimizing, presutti2020focusing}. This interplay highlights the inherent complexity in optimizing these lenses, where improvements in one aspect may lead to compromises in others. In addition, meta-atom-based metalenses exhibit narrow field of view (FoV) stemming from angular dispersion inherent in meta-atom-based designs~\cite{so2022revisiting}. Consequently, at present, reported broadband metalenses exhibit chromatic aberration~\cite{presutti2020focusing,fan2020ultrawide} or low focusing efficiency over a large bandwidth~\cite{wang2018broadband, shrestha2018broadband}, which impedes the commercialization of metalens-based compact imaging. This compromise, by rendering the attainment of high-efficiency broadband focusing alongside minimal chromatic and angular aberration a considerable challenge, substantially restricts the performance and the range of potential applications of metalenses. Even for ideal metalens, it may not simultaneously satisfy broadband operation and large diameter due to the physical upper bounds~\cite{miller2021fundamental}. Moreover, the limitations inherent in conventional design approaches complicate efforts to effectively address these challenges in metalens development. 

In  direct response to these challenges, we introduce an innovative, deep-learning-powered, end-to-end integrated imaging system. By synergizing a specially designed large-area mass-produced metalens~\cite{kim2023scalable} with customized image restoration framework, we propose a comprehensive imaging solution poised to supplant conventional geometric lens-based systems. The proposed system  not only effectively addresses the aberrations mentioned above but also leverages the inherent strengths of large-area mass-produced metalenses to make a significant step toward high-quality, aberration-free images. Moreover, our approach distinguishes itself by suggesting a metalens image restoration framework that can fit any metalenses suffering from aberrations or low efficiency. Also, assuming the uniform quality of mass-produced metalenses, the optimized restoration model can be applied to other metalenses manufactured at the same process. The proposed imaging system may pave the way for the next generation of compact, efficient, and commercially viable imaging systems.

In summary, this work propels metalens technology to new heights and underscores the transformative potential of deep learning in initiating a paradigm shift in optical imaging. Through our end-to-end imaging framework, we not only demonstrate a viable pathway to surmount traditional optical limitations but also pave the way for a novel era in compact and efficient imaging solutions.   This breakthrough has the potential to revolutionize the field of optical engineering, sparking new avenues of research and innovation.

\section{Methods}

\begin{figure}
    \centering
    \includegraphics[width=1.0\linewidth]{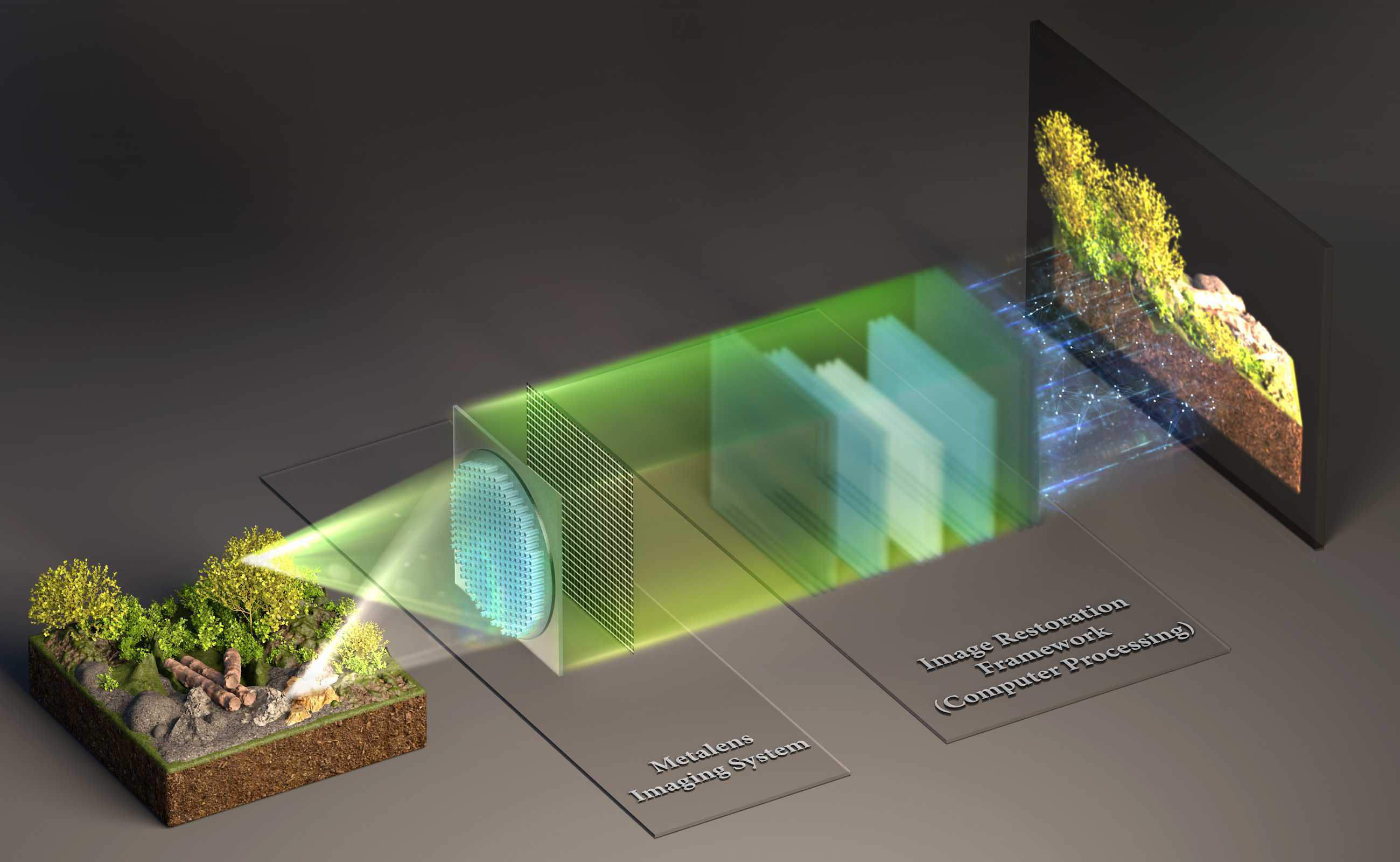}
    
    \caption {Schematic of our metalens imaging.}
    \label{fig:fig1}
\end{figure}

A schematic of our end-to-end integrated imaging system is shown in Fig.~\ref{fig:fig1}. This system combines a metalens-based imaging system and a subsequent image restoration framework. The former component is tasked with acquiring the image, while the latter is responsible for restoring the captured image. When tailored to restore the image produced by the metalens imaging system automatically, the framework can independently generate an output image that closely approximates the quality of the ground truth image.

The metalens designed in this work is composed of an array of nanostructures with arbitrary rotational angles, the class of metalenses designed this way being known as the Pancharatnam–Berry (PB) phase-based metalens. Despite the ability of these PB-phase-based metalenses to achieve diffraction-limited focusing~\cite{khorasaninejad2016metalenses, kim2023scalable}, they are not without their challenges. The dispersion of the meta-atoms can induce chromatic aberration, a characteristic similarly observed in diffractive lenses~\cite{engelberg2017optimizing}. Substantial efforts have been made to achieve achromatic metalenses through dispersion engineering of meta-atoms~\cite{shrestha2018broadband, wang2018broadband}, adjoint optimization~\cite{chung2020high, lin2019overlapping}, and many other methods~\cite{huang2020design, khorasaninejad2017achromatic}. However, the resulting metalenses still suffer from relatively low efficiency compared to single-frequency metalenses. Also, the PB-phase-based metalenses are concurrently susceptible to angular aberration that originates both from the Seidel aberrations~\cite{yang2023wide} and angular dispersion of the meta-atom~\cite{so2022revisiting}.  The combination of these factors sets our full color high resolution imaging apart from conventional restoration tasks~\cite{NAFNet, Zamir2021Restormer}, thereby significantly complicating the task of restoring images captured by the metalens to their original state. Our framework thus addresses and rectifies the aberration issues of the metalens using a customized deep learning approach.

Specifically, prior to training, we gathered hundreds of aberrant images captured by the metalens imaging system, which we refer to as ``metalens images". The metalens images, which exhibit the physical defects of the metalens, were then used to train the image restoration framework. The result is a significant enhancement in the quality of the image produced by the compact metalens imaging system. The framework employed in this process is composed of two primary stages. In the first stage, the framework is optimized to reduce the discrepancy between the outputs of its restoration model and the ground truth images. Following this, an adversarial learning scheme that incorporates an auxiliary discriminator is utilized to augment the image restoration model’s ability to recover lost information.

By concatenating our restoration framework to our imaging system comprised of our mass-produced metalens, we construct an integrated imaging system that delivers high-quality compact imaging. This system is scalable to larger apertures and different wavelengths, thereby offering an optimal solution for a novel miniaturized imaging scheme. Importantly, the reproducibility of both the imaging system and the restoration framework not only enhances the commercial viability of this integrated system but also suggests that the commercial application of metalenses could become a reality in the near future. In the following subsections, we elaborate on the construction of the integrated system, starting from the metalens to the image restoration framework. 

\subsection{Metalens Imaging System}

\begin{figure}
    \centering
    \includegraphics[width=1.0\linewidth]{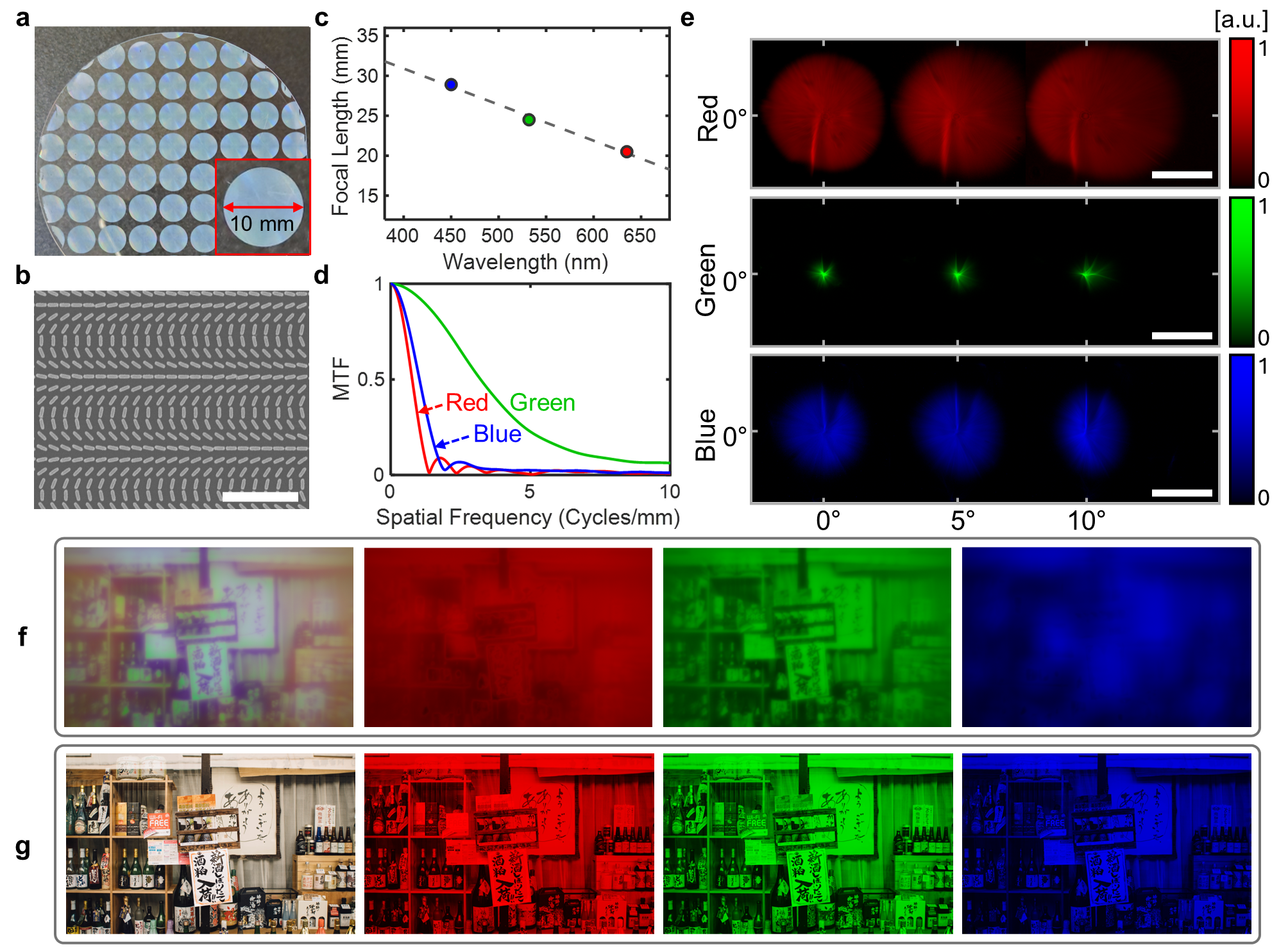}
    \caption {(a) Photographs of fabricated mass-produced 10-mm-diameter metalenses on \ang{;;4} glass wafer. The inset in the red box shows enlarged image of the metalens. (b) Scanning Electron Microscopy (SEM) image showing the meta-atoms that compose the metalens. The scalebar is 3 \unit{\micro\metre}. (c) Focal lengths of the metalens for wavelengths of 450 (blue), 532 (green), and 635 (red) nm. The dashed line indicates the linear fitting result. (d) MTFs of red, green, and blue lights with zero viewing angle. (e) PSFs of red, green and blue lights with various viewing angles (\ang{0}, \ang{5}, \ang{10}). The scale bar is 1mm, which indicates a distance on the image sensor. (f) The metalens image (left) and its subset images showing red, green, and blue color channels. (g) Corresponding ground truth image (left) and its subset images showing red, green, and blue color channels.} 
    \label{fig:fig2}
\end{figure}
The metalenses are fabricated through nanoimprint lithography and subsequent atomic layer deposition~\cite{kim2023scalable}. The nanoimprint lithography provides the benefits of low-cost mass-production and uniformity of the products.~\cite{kim2023scalable, kim2023one, kim2022metasurface, yoon2021printable} Thus, we use the imprinted metalenses to broadly impact our work on the commercialization of the DNN-based metalens imaging system. Figure~\ref{fig:fig2}(a) shows the mass-produced 10-mm-diameter metalenses fabricated by nanoimprint lithography and subsequent thin-film deposition of TiO$_2$, the details of the fabrication process in the Supplementary material. As shown in Fig.~\ref{fig:fig2}(b), our metalens comprises nano-slabs with arbitrary rotational angles as a PB-phase-based metalens. It has a relatively high efficiency of 55.6\% at 532 nm wavelength but exhibits severe chromatic aberration. As shown in Fig.~\ref{fig:fig2}(c), the focal lengths at wavelengths of 450 nm, 532 nm, and 635 nm are 29.0, 24.5, and 20.5 mm, respectively~\cite{kim2023scalable}.  This wavelength-dependent focal length results in Transverse Axial Chromatic aberration (TAC), which can be expressed as\begin{equation} 
    \text{TAC} = |f - f_0| \frac{D}{2f} 
\end{equation} where $f$ is the focal length of the incident light with a single wavelength; $f_0$ is the distance between the metalens and image sensor; and $D$ is the diameter of the metalens. When $f$ differs from $f_0$, the incident light forms a top-hat-like PSF profile with a radius equal to that of TAC~\cite{engelberg2017optimizing}. Given this, the reduction in the overall TAC in the visible band is important for high-quality metalens imaging because the blur of the image intensifies proportionally with the increase in TAC. Thus, we chose $f_0$ as the focal length of green light (24.5 mm) to minimize the overall TAC. As a result, the red (635 nm) and blue (450 nm) lights are defocused than green light with the TACs of 0.98 and 0.78 mm, respectively.

The metalens imaging system is affected by chromatic and angular aberrations and its surface defects by incomplete fabrication. To quantify these effects, we measured the Point Spread Functions (PSFs) and calculated the Modulation Transfer Function (MTF) from the measured PSFs. The PSF, which is the 2D intensity distribution obtained in response to a single point light source~\cite{goodman2005introduction}, is a critical metric for evaluating the quality of an imaging system because it is directly related to image formation~\cite{zhou2021image}. The MTF, calculated using the measured PSFs, describes the imaging quality in terms of resolution and contrast\cite{goodman2005introduction}. We measured the PSFs by capturing the images of red, green, and blue collimated beams using the metalens imaging system, and subsequently calculated the MTFs from the PSFs. The PSF measurement and imaging setup and the MTF calculation method are in supplementary information.

Figure~\ref{fig:fig2}(e) shows the PSF of red, green, blue lights with various viewing angles (\ang{0}, \ang{5}, \ang{10}). The PSF profile profiles of red and green lights show wide disk shapes while the profile of the green light shows irregular spark shape, implying the effect of TAC. Thus, as shown in Fig.~\ref{fig:fig2}(d), the MTFs of the red and blue lights are severely lower than the MTF of the green light at all spatial frequencies. Furthermore, the PSF profiles at zero viewing angle show non-ideal and circularly asymmetric shapes, which can be attributed by the defects of the metalens due to the imperfect fabrication. The PSF profiles also changes their shapes with viewing angle due to the angular aberrations including Seidel aberrations~\cite{yang2023wide} and the angular dispersion of the meta-atoms~\cite{so2022revisiting}. The PSF profiles of the red and green lights stretches to the horizontal direction as the viewing angle increases, where the profile of the blue light shrinks. In addition, the non-uniformity of the metasurface during the fabrication process~\cite{kim2023scalable} may result in the PSFs with complex profiles not matching the PSFs from the Rayleigh-Sommerfeld diffraction formula~\cite{shen2006fast}. As a result, the combination of these generates the complex PSF profiles varying with the viewing angle and further complicates the image restoration tasks. 

The effects of chromatic and angular aberrations to the metalens images can be shown by comparing the image with the ground truth image. Figures~\ref{fig:fig2}(f) and (g) show the metalens image, the corresponding ground truth image, and the subset images depicting the red, green, and blue color channels. The red and blue channels of the metalens image are severely blurred from TAC, making it difficult to recognize any objects. The green channel of the metalens image shows a relatively higher resolution at the center, which gradually decreases as the viewing angle increases (e.g., the outer region of the image) due to the angular aberrations at the higher viewing angle. 

\subsection{Image Restoration Network}

Computational image restorations have emerged as a prevalent approach for the enhancement of non-ideal images, such as those that are noisy~\cite{li2023ntire} or blurred~\cite{nah2021ntire}. Classical image restoration methods achieve higher resolution by relying on linear deconvolution methods, such as applying the Wiener filter~\cite{goldstein1998multistage}. Deconvolution, an inverse of the convolution operation, facilitates the recovery of the original image from an image convolved with a PSF. The performance of the deconvolution process depends on two factors: the space invariance of PSF across the FOV and the low condition number for the inverse of the PSF~\cite{heath2018scientific}. However, Wiener filters exhibit limited restoration quality for imaging systems with PSFs that vary depending on the viewing angle, such as metalens imaging systems~\cite{huang2020design} and under display cameras~\cite{zhou2021image}.

An alternative restoration approach is the utilization of DNN-based image restoration. DNN-based restoration models~\cite{NAFNet, Zamir2021Restormer} have shown superior performance compared to traditional approaches in specialized tasks such as denoising~\cite{li2023ntire}, de-blurring~\cite{nah2021ntire}, super-resolution~\cite{yang2022ntire}, and light-enhancement~\cite{wu2022uretinex}. Furthermore, they are applicable to imaging systems with complex and combined degradations such as under-display cameras~\cite{zhou2021image} and the \ang{360} FOV panoramic camera~\cite{jiang2022annular}. However, conventional DNN approaches are incapable of learning position-variant image degradations (e.g., position-dependent aberration of the metalens) because these methods train models with randomly cropped patches from full-resolution images, leading to the complete loss of position-dependent information.

In response to these challenges, we propose an end-to-end image restoration framework specifically tailored for the metalens imaging system to address non-uniform aberration over wavelength and viewing angle. Contrary to the images that are subjected to restoration in typical image restoration tasks~\cite{Agustsson_2017_CVPR_Workshops, Nah_2017_CVPR}, our metalens images exhibit more intense blur and significant color distortion. Consequently, the restoration of metalens images constitutes a severely ill-posed inverse problem. To address this critically underconstrained problem, we employ strong regularization. That is, we model the traits and patterns of sharp data, performing adversarial learning in Fourier space to train the data distribution. Therefore, the restoration model $f(y)$ is trained by minimizing 

\begin{equation}  \label{eq:2} 
    L(x, y, f) = E(x, f(y)) + \lambda \Phi(f(y)) 
\end{equation} where $E(x, f(y))$ is the image fidelity term that approximates the restored metalens image $f(y)$ by the ground truth image $x$, and $\Phi(f(y))$ is the regularization term that limits the space of $f(y)$. Subsequently, we apply positional embedding to learn the angular aberration of metalens imaging. Because the proposed method utilizes information on the absolute coordinates of randomly cropped patches, the model effectively trains the highly space-variant degradations. 

\subsubsection{Network Architecture}
\begin{figure}
    \centering
    \includegraphics[width=1.0\linewidth]{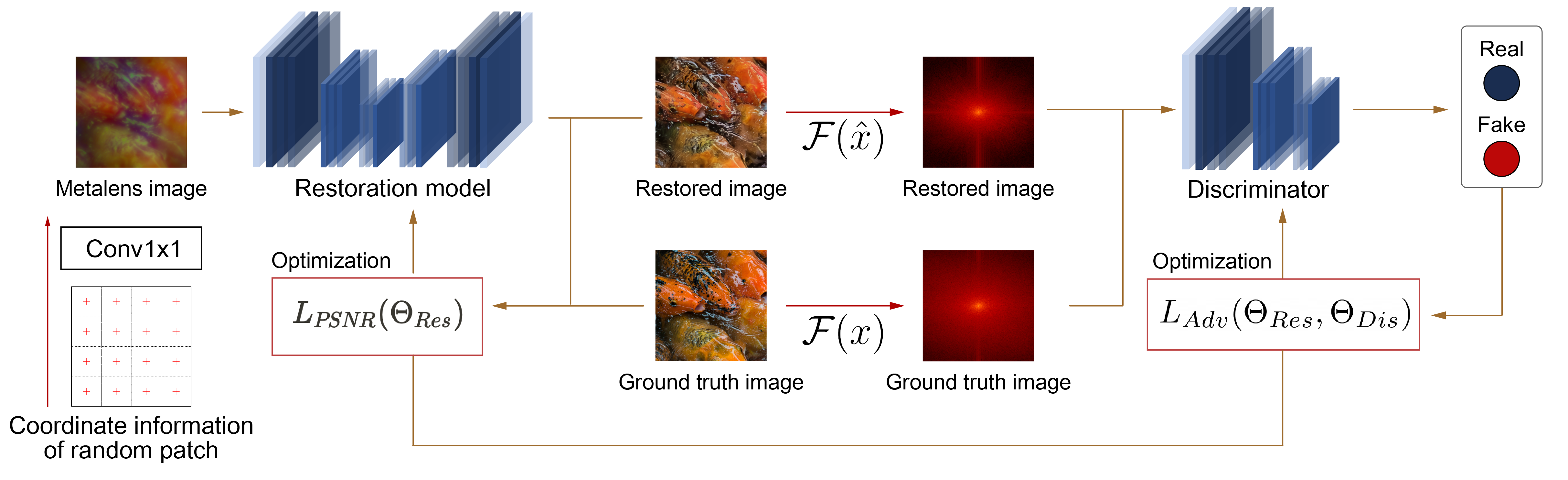}
    \caption {Proposed image restoration framework. The framework consists of an image restoration model, and applies random cropping and position embedding to the input data using coordinate information of the cropped patches. To address the underconstrained problem of restoring degraded images to latent sharp images, adversarial learning in the frequency domain is applied through the fast Fourier transform(F). $\hat{x}$ and $x$ denote the reconstructed and ground truth image, respectively. The details of the framework are in the Methods section.} 
    \label{fig:fig3}
\end{figure}

The architecture of our image restoration framework is depicted in Fig.~\ref{fig:fig3}. Our framework incorporates the existing DNN architecture with our proposed methods. The training phase involved the utilization of patches randomly cropped from images at their full resolution, specifically $1280\times800$ in this study. Subsequently, in the inference phase, the analysis was conducted on the entire images at their original resolution of $1280\times800$. However, we observed a statistical disagreement in the inference process of the full-resolution image, as shown in Fig.~S3. To overcome this inconsistency, we apply test-time local converter (TLC)~\cite{chu2022improving} in test phase, which yields a significant performance improvement. The detailed results are presented in Table~S1.

The metalens used in our study exhibits intense chromatic and angular aberrations, resulting in severe information loss in the images captured with it. Therefore, we trained the model according to the traits and patterns found in the underlying clean images to efficiently restore a wide range of spatial-frequencies and constrain the space of the latent ground truth images. Because generative models can learn complex, high-dimensional data distributions from a given dataset~\cite{ruthotto2021introduction}, we utilized an adversarial learning scheme, one of the generative learning methods, to learn effectively the distribution of latent sharp images by introducing an auxiliary discriminator. We initially applied adversarial learning in the RGB space but found that conspicuous pattern artifacts occurred in both RGB and Fourier spaces (Fig.~S4). Therefore, we transformed the data from each channel’s RGB space into Fourier space, a process which prevents the occurrence of pattern artifacts while restoring high-frequency information. These Fourier space data are then used as the input data for the discriminator.

The training loss is composed of two distinct terms: peak signal-to-noise ratio (PSNR) loss $L_{\mathrm{PSNR}}$ between the reconstructed image $\hat{x}$ and the ground truth image $x$, and adversarial loss $L_{a}$ between $\mathcal{F}(\hat x)$ and $\mathcal{F}(x)$. The PSNR loss signifies the image fidelity loss, and the adversarial loss indicates the prior regularization loss. Therefore, the total loss function $L_{\mathrm{Total}}$ is: 

\begin{equation} 
    L_{\mathrm{Total}}= L_{\mathrm{PSNR}} + \lambda L_{\mathrm{a}}  
\end{equation}
where $\lambda$ is a hyperparameter for balancing $L_{\mathrm{PSNR}}$ and $L_{\mathrm{a}}$. $L_{\mathrm{PSNR}}$ is calculated as follows: 

\begin{equation} 
    L_{\mathrm{PSNR}}(\hat{x}, x) = -10\log {R^2 \above 1pt \mathrm{MSE}(\hat{x}, x)} 
\end{equation}
where $\hat{x}, x$ and $R$ denote the reconstructed image, ground truth image, and the maximum signal value of the ground truth image, respectively. MSE is the distance between the reconstructed and ground truth images and is formulated as $\mathrm{MSE}(\hat{x}, x) = {1\above 1pt N}\sum_{n=1}^N (\hat{x}_n - x_n)^2$.

For adversarial learning, we constructed an additional discriminator and applied spectral normalization~\cite{miyato2018spectral} for training stability. In addition, we employ the GAN training scheme based on hinge loss~\cite{lim2017geometric} for enhanced stability of adversarial training. The adversarial loss $L_{a}$ of the discriminator (D) and the image restoration model (G) is:  

\begin{equation} 
L^\mathrm{D}_a = \mathbb{E}_{x}[\max(0, 1-D(\mathcal{F}(x)))] + \mathbb{E}_{\hat{x}}[\max (0, 1+D( \mathcal{F}(\hat{x})))] 
\end{equation} 

\begin{equation} 
L^\mathrm{G}_a = -\mathbb{E}_{\hat{x}}[D(\mathcal{F}(\hat{x}))]   
\end{equation} where $\mathcal{F}$ refers to FFT. Here $\mathbb{E}_{x}[\cdot]$ and $\mathbb{E}_{\hat{x}}[\cdot]$ are operators that denote the calculation of the mean of the ground truth and reconstructed images in the given minibatch, respectively. The image restoration model and discriminator each try to minimize $L^\mathrm{G}_a$ and $L^\mathrm{D}_a$, respectively.

Degradation in the outer region of the metalens image is more pronounced than in the central region due to the angular aberration. This observation suggests that positional information is integral for understanding the degradation of the metalens imaging system. However, the training method makes it impossible for the model to learn positional information because our framework learns through random patches during training and restores full-resolution images during inference.

To address this problem, we took the coordinate values of each pixel of the patches, based on the coordinates of a full resolution image, and map them through a $1\times1$ convolutional layer. This process transforms them into proper space when generating random patches for a full-resolution image. The processed coordinate information is concatenated with metalens images corresponding to the information. The resulting concatenated data is used as input data. This approach enables the model to learn and leverage positional information effectively, thereby enhancing its performance in restoring full-resolution images. 

\subsubsection{Data Acquisition}
The training data for the metalens imaging system was obtained by capturing ground truth images displayed on the \ang{;;85} monitor~Fig.~S1. For training, we utilized the DIV2K dataset~\cite{Agustsson_2017_CVPR_Workshops}. This dataset contains 2K resolution images of various objects, thereby providing environmental diversity. The ground truth images for training were obtained by cropping the center of the dataset images by 1280×800 resolution to ensure that the ground truth images fit within the Field of View (FOV) of the metalens imaging system.

The positions of the objects in both the metalens image and the corresponding ground truth image were matched for effective training. Raw metalens images with $5472\times3648$ resolution were rotated, cropped to $5328\times3328$ resolution, and resized to $1280\times800$ resolution to match the corresponding ground truth images. The rotation angle and cropping parameters were optimized to maximize the SSIM between the metalens images and the corresponding ground truth images. Finally, we divide the dataset into 628 and 70 images for training and testing, respectively. 

\subsubsection{Training Details}
As mentioned in network architecture section, training was conducted using patches that were randomly cropped from full resolution images. While larger receptive fields offer more comprehensive semantic information, they also increase the training time and computational complexity. Consequently, to strike a balance between performance and training duration in the proposed model, we set the patch size at 256$\times$256 and the batch size at 16. In addition, transformations such as horizontal and vertical flips and transpositions were randomly applied, then coordinate information of the patches were loaded under these configurations.

The model used in this paper can be divided into two components; the first of which is the image restoration model. The width of the starting layer of the network is set to 32, which doubles as the network delves deeper into each successive block. The encoder and decoder of the network are each composed of four blocks. To address the inconsistency between training and testing, Test-time Local Converter (TLC) is adopted during the testing phase. The number of input and output channels of $1\times 1$ convolutional layer which processes coordinate information are both set to 2. The second part is the discriminator, where its width is set to 64, and all layers had the same width. The discriminator is composed of five blocks. Moreover, spectral normalization~\cite{miyato2018spectral} is applied to stabilize the learning process.

The training was executed in two stages. In the first stage, the metalens images were restored to clean images using the image restoration model, and in the second stage adversarial learning was performed using the discriminator after expressing the restored and ground truth images in the spatial frequency domain through fast Fourier transform (FFT). Because the spatial-frequency domain data converted through FFT are complex (comprising real and imaginary parts), these parts were represented as a two-dimensional vector. This allowed the data in the spatial-frequency domain to be expressed as real vectors, which were then used as input into the discriminator.

During the training process, the number of iterations was set to 300,000. In the image restoration model, AdamW was used as the optimizer with the learning rate initially set to $3 \times 10^{-4}$ and gradually decreased to $10^{-7}$ following the cosine annealing schedule; the betas were $[0.9, 0.9]$. For the discriminator optimizer, Adam was used with the learning rate set to $3\times10^{-4}$, identical to the restoration model, but the betas were set to $[0.0, 0.9]$; NVIDIA RTX 4090 24GB was used as a computational resource for this training. 

\subsubsection{Statistics Details}
Statistical hypothesis testing was performed using the statistical functions of the SciPy library in Python using 70 test images. The two-sided paired t-tests were used to compare the performances of models at the $P < 10^{-4}$ significance level.

\section{Results}

In this study, we have introduced a deep-learning-powered, end-to-end integrated imaging system. We now assess its capability in various perspectives to restore metalens images to their clean states, addressing severe chromatic and angular aberrations inherent in our large-area mass-produced metalens. In order to draw a comparison between the images produced by our framework and those captured with the metalens, we restored a total of 70 metalens images to their undistorted state. Given these pairs of images, we conduct a thorough assessment of our system’s efficacy in image restoration, employing a comprehensive set of performance metrics tailored to each category of interest under evaluation. We also compare our framework with state-of-the-art models, including restoration models for natural images (MIRNetv2~\cite{Zamir2022MIRNetv2}, HINet~\cite{Chen_2021_CVPR}, NAFNet~\cite{NAFNet}).  

\begin{figure}
    \centering
    \includegraphics[width=1.0\linewidth]{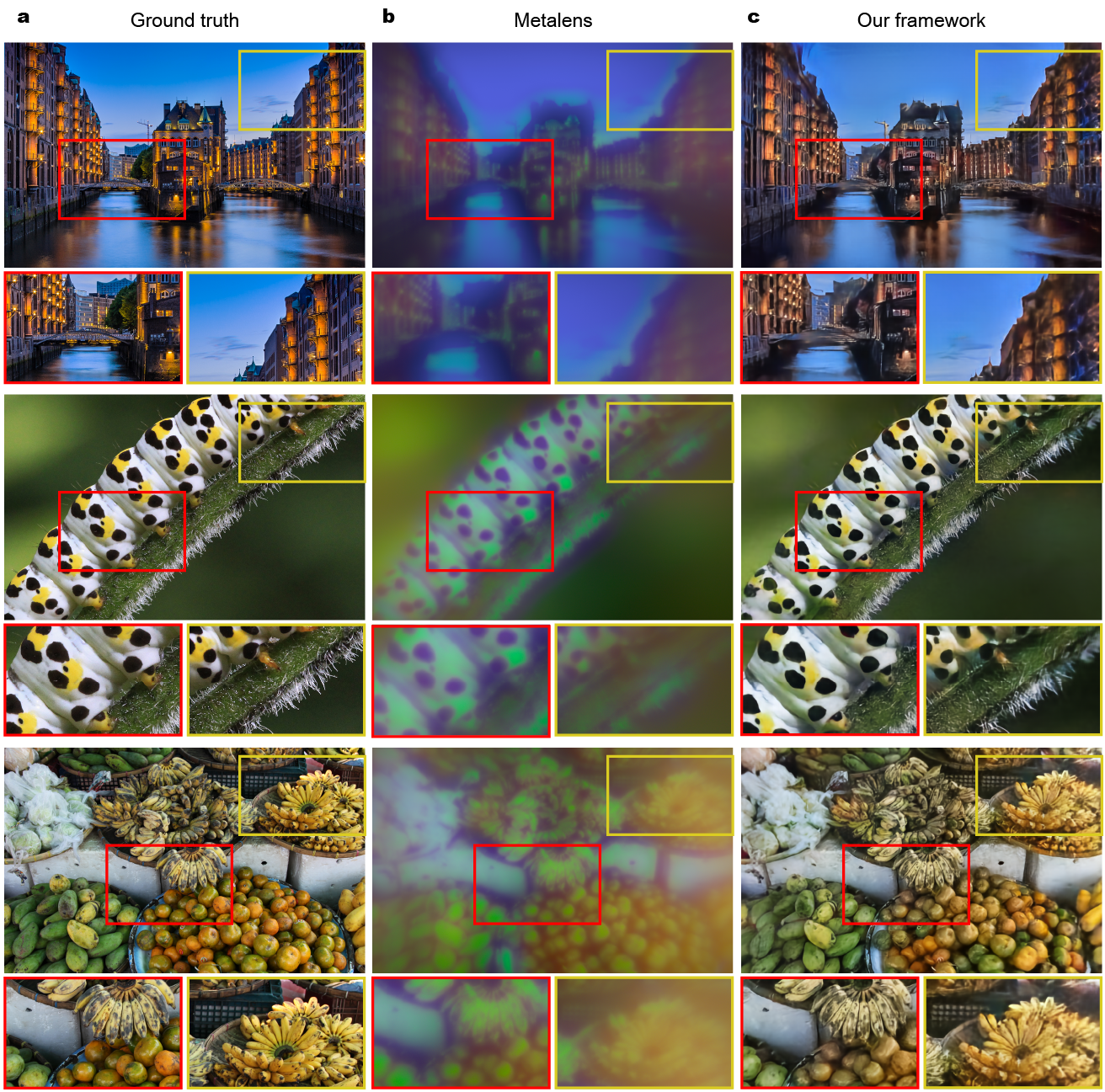}
    \caption {(a) Ground truth images, (b) metalens images, and (c) images reconstructed by our model. The images are affiliated with the test set data. The central (red) and outer (yellow) regions of the images are enlarged to access the restoration of the metalens image at high and low viewing angle, respectively. The outer regions of the metalens images (yellow box) are successfully restored, even though those are more severely degraded than the inner region (red box) due to the angular aberration under high viewing angle. } 
    \label{fig:fig4}
\end{figure}

Figure~\ref{fig:fig4} comprehensively shows the qualitative restoration results of our integrated imaging system by comparing the ground truth, metalens, and the system outcome images. Notably, the images captured by the metalens are marred by pronounced chromatic aberrations, manifesting as a noticeable disparity in the clarity of red and blue components in comparison to green, thereby engendering significant blurring. Furthermore, this aberration is accompanied by a loss in high-frequency information, leading to the erosion of fine details present in the original images. A particularly marked manifestation of this degradation is observed in the peripheral regions (marked by a yellow box) as compared to the central zone (highlighted by a red box) in Fig.~\ref{fig:fig4}, where the images exhibit enhanced blurring, resulting in the obliteration of sharp details and the predominance of a specific hue. 

Contrastingly, images reconstructed utilizing our proposed framework exhibit a remarkable fidelity to the ground truth across both peripheral and central regions, demonstrating the framework's proficiency in reinstating details obliterated by chromatic aberration. Such outcomes underscore the capability of our framework to surmount the intricate challenges posed by a highly irregular Point Spread Function (PSF), thereby significantly augmenting the imaging performance across a spectrum of scenarios. This denotes a substantial stride towards mitigating the complexities associated with aberration-induced degradation, heralding advancements in the fidelity and quality of imaging systems employing metalenses. 

Despite the physical limitations inherent in metalenses, which cannot be overcome through conventional manufacturing processes alone, our application of deep learning enables imaging capabilities that exceed the physical performance limits of the metalenses. This innovative approach effectively bridges the gap between the inherent physical constraints and the desired imaging outcomes. 

In the following subsections, we present a comparative statistical analysis based on the test dataset to assess the quality of image restoration. This analysis further illustrates how our deep learning-enhanced framework not only compensates for the physical limitations of metalenses but also significantly improves the overall image quality.

\subsection{Quality of Image Restoration}

\begin{figure}
    \centering
    \includegraphics[width=1.0\linewidth]{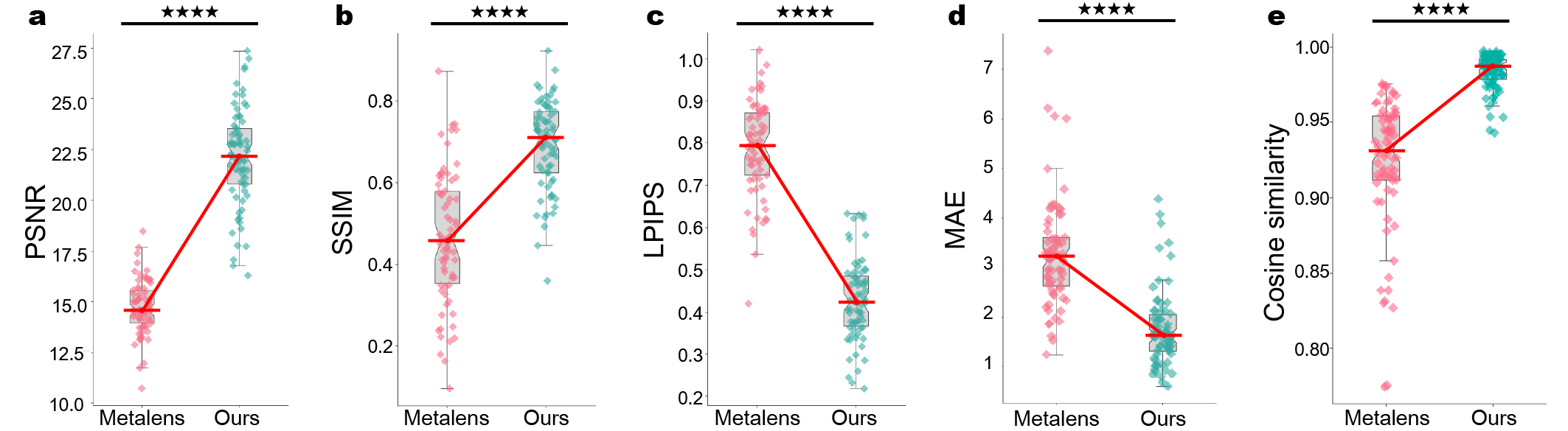}
    \caption {Comparative statistical analysis of the proposed model and metalens imaging results using the test dataset. (a-e) Results of PSNR, SSIM, LPIPS in RGB space and cosine similarity, MAE of the magnitudes in Fourier space calculated by comparing the metalens image and the image reconstructed by our framework with the ground truth image. A statistical hypothesis test was performed through a two-sided paired t-test on the performance difference between the metalens image and the image reconstructed by our framework (significance level $P = 10^{-4}$, (a) $1.055*10^{-39}$,  (b) $3.886 * 10^{-35}$, (c) $1.363*10^{-48}$, (d) $2.311*10^{-35}$, (e) $2.150*10^{-38}$).} 
    \label{fig:fig5}
\end{figure}

\hspace{\parindent}
Figure~\ref{fig:fig5} comprehensively shows the results of PSNR, structural similarity index measure (SSIM), learned perceptual image patch similarity (LPIPS) in RGB space and mean absolute error (MAE) of the magnitudes, cosine similarity (CS) in Fourier space calculated by comparing the metalens image and the image reconstructed by our framework with the ground truth image. The red horizontal lines in each box represent the median, and the boxes extend from the first to the third quartile. The whiskers span 1.5 times the interquartile range of the first and third quartiles. We conducted a statistical hypothesis test to ascertain whether the observed results exhibit statistically significant differences. This was accomplished through the utilization of a two-sided paired t-test to evaluate the performance disparity between images produced by metalenses and those reconstructed by the proposed framework. A significance level of $P = 10^{-4}$ was set for the testing process. 

Within this analysis, the outcomes indicate a statistically significant variance across all evaluated metrics, as evidenced in Fig.~\ref{fig:fig5}. These metrics were assessed utilizing a test set comprising 70 data points. Also, Table \ref{tab:Tab1} shows the quantitative results of the metalens imaging system, our framework, and state-of-the-art models for various metrics. The implications derived from each graph and the significance of the quantitative outcomes are elaborated below, providing a comprehensive analysis of the data and its relevance to the study’s objectives.

\begin{table}[h]
    \centering
    {\begin{tabular}{cccccccc}
    \toprule
    \multicolumn{1}{c}{} & \multicolumn{3}{c}{\textbf{Image quality metric}} & \multicolumn{2}{c}{\textbf{Assessment in frequency domain}} \\
    \cmidrule(rl){2-4} \cmidrule(rl){5-6}
    \textbf{Model} & {PSNR} & {SSIM} & {LPIPS} & {MAE} & {CS}\\
    \midrule
    Metalens image & 14.722/1.328 & 0.464/0.160 & 0.788/0.112 & 3.281/1.089 & 0.922/0.045\\
    MIRNetV2 & 18.507/1.893 & 0.591/0.135 & 0.559/0.098 & 2.240/0.900 & 0.967/0.020\\
    SFNet & 18.223/1.727 & 0.602/0.129 & 0.519/0.095 & 2.194/0.837 & 0.965/0.020\\
    HINet & 21.364/2.333 & 0.677/0.117 & 0.456/0.097 & 1.851/0.800 & 0.982/0.013\\
    NAFNet & 21.689/2.382 & 0.676/0.116 & 0.440/0.097 & 1.817/0.801 & 0.983/0.013\\
    \rowcolor{lavender}
    Our framework & \textcolor{red}{\textbf{22.095}}/2.423 & \textcolor{red}{\textbf{0.692}}/0.109 & \textcolor{red}{\textbf{0.432}}/0.096 & \textcolor{red}{\textbf{1.759}}/0.779 & \textcolor{red}{\textbf{0.984}}/0.012\\
    \bottomrule
    \end{tabular}}
    \caption{\label{tab:Tab1} Comparison of quantitative assessments of various models using the test set of images (n=70). The first and second values of each column represent the mean and the standard derivation of the metrics, respectively. The best scores are marked as red and bold.}
\end{table}

To further understand the impact of our framework on the fidelity of image restoration, we examine PSNR and SSIM~\cite{wang2004image}, which serve as foundational metrics. The former is a quantitative measure of the restoration quality of an image, calculated as the logarithmic ratio between the maximum possible power of a signal (image) and the power of corrupting noise that affects its fidelity. Higher PSNR values indicate better quality of the reconstructed image. The latter, SSIM, valuates the visual impact of three characteristics of an image: luminance, contrast, and structure, thus providing a more accurate reflection of perceived image quality.

Figure~\ref{fig:fig5} presents a statistical analysis comparing the PSNR and SSIM values of the images captured through the metalens with those restored by our framework. As shown in Table \ref{tab:Tab1}, the framework showcased a remarkable improvement in image fidelity, elevating the PSNR by 7.37 dB and SSIM by 22.8\%p compared to the original metalens images. These enhancements underscore our framework's proficiency in mitigating the fidelity loss incurred by metalens aberrations, thus significantly elevating the quality of the reconstructed images closer to their ground truths.

While PSNR and SSIM are advantageous for assessing image fidelity and perceived quality, they often fall short in evaluating the structured outputs. This limitation stems from their inability to fully capture the human visual system's sensitivity to various image distortions, particularly in textured or detailed regions. To address this gap, LPIPS~\cite{zhang2018unreasonable} was employed to evaluate the perceptual quality of the images. LPIPS evaluates perceptual similarity by utilizing pretrained deep learning networks (e.g. AlexNet), offering a nuanced measure that aligns more closely with human perception of image quality. Lower LPIPS values indicate better perceptual quality.

Table \ref{tab:Tab1} demonstrates that our framework achieved a 35.6\%p decrease in LPIPS, indicating a substantial enhancement in the perceptual resemblance of the reconstructed images to their original counterparts, as also observable in Fig.~\ref{fig:fig5}(c). This metric highlights the proposed framework's capability to not only improve the objective quality of images but also their subjective, perceptual quality.We also compare our framework with state-of-the-art models, including restoration models for natural images ( MIRNetv2~\cite{Zamir2022MIRNetv2}, HINet~\cite{Chen_2021_CVPR}, NAFNet~\cite{NAFNet}). As shown in Table~\ref{tab:Tab1}, our framework surpasses these state-of-the-art models by a substantial margin in terms of PSNR, SSIM, and LPIPS. This suggests that our framework is more suitable for the metalens image restoration task than conventional models designed for restoring natural images, such as those in the DIV2K dataset~\cite{Agustsson_2017_CVPR_Workshops}.

The measured MTF of the metalens in Fig.~\ref{fig:fig1}(d) and qualitative results in Fig.~\ref{fig:fig4}(b) demonstrate intense degradation at high spatial frequencies. Consequently, it is crucial to restore the spatial-frequency information during the metalens image restoration task. It is pertinent to acknowledge that spatial frequency can be represented as both magnitude and phase components, with the latter often heralded as important in signal processing realms ~\cite{oppenheim1981importance}. We utilize two metrics to evaluate the magnitude and phase of the Fourier-transformed reconstructed images. In evaluating the fidelity of the reconstructed images, particularly concerning their frequency-dependent attributes, two metrics are employed: the Mean Absolute Error (MAE) for assessing discrepancies in magnitude relative to the original images, and the Cosine Similarity (CS) for gauging phase congruence with the authentic images. These metrics are derived through the application of the Fast Fourier Transform (FFT) across images revitalized by disparate models. The ensuing MAE and CS metrics underscore a remarkable enhancement in image quality, as elucidated in Fig.~\ref{fig:fig5}(d-e) and Table~\ref{tab:Tab1}. As shown in these figures, our framework demonstrates the superior performance of MAE and CS to the metalens imaging system and several state-of-the-art image restoration models in the frequency domain. Our framework achieved about twice the performance of the metalens imaging system for MAE and accomplished about 14\%p for CS for the metalens images.

\begin{figure}
    \centering
    \includegraphics[width=1.0\linewidth]{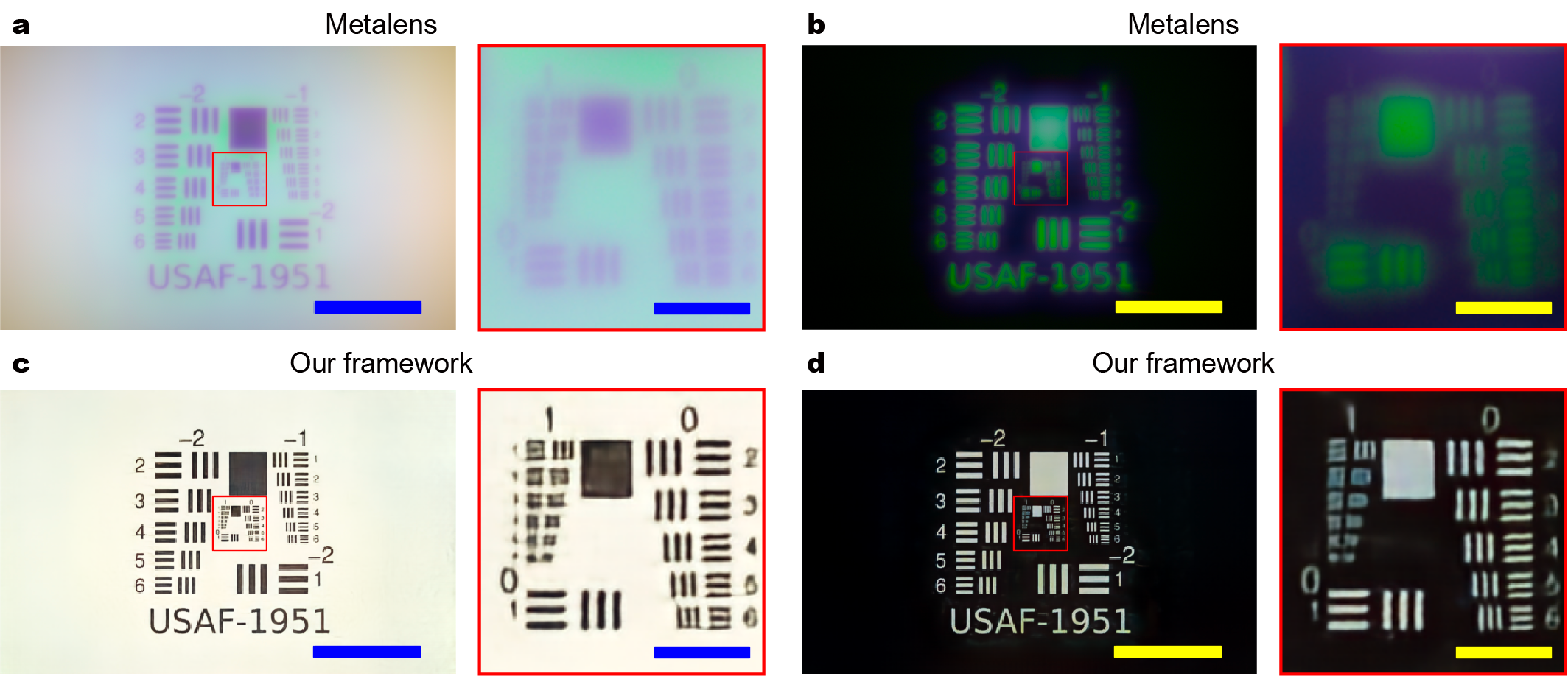}
    \caption {(a) and (b) White and black USAF images captured by the metalens imaging system, respectively. (c) and (d) White and black USAF images restored by our framework, respectively. The image in the red boxes shows the enlarged images at the central region indicated as red box. The scalebars in the original and enlarged images are 3 and 0.5 mm, respectively, indicating the distance on the image sensor.} 
    \label{fig:fig6}
\end{figure}

To demonstrate the restoration of the blur and color distortion visually, we tested our imaging system using 1951 U.S. Air Force resolution test chart images (USAF images). Figures~\ref{fig:fig6}(a-b) show monochromatic white and black USAF images captured by the metalens imaging system. These images exhibit severe blurring and strong color distortion, particularly showing greenish tints in white patterns. As shown in Fig.~\ref{fig:fig6}(c) and (d), the restored images illustrate the pattern’s colors are closer to white and black than the metalens images. Furthermore, the central regions of the images exhibit high sharpness, while the damaged images have severe blurring in these areas. Thus, our framework demonstrates superior prominence in enhancing overall image quality by achieving conspicuous color fidelity and sharpness. 

\subsection{Object Detection Performance}

\begin{figure}
    \centering
    \includegraphics[width=1.0\linewidth]{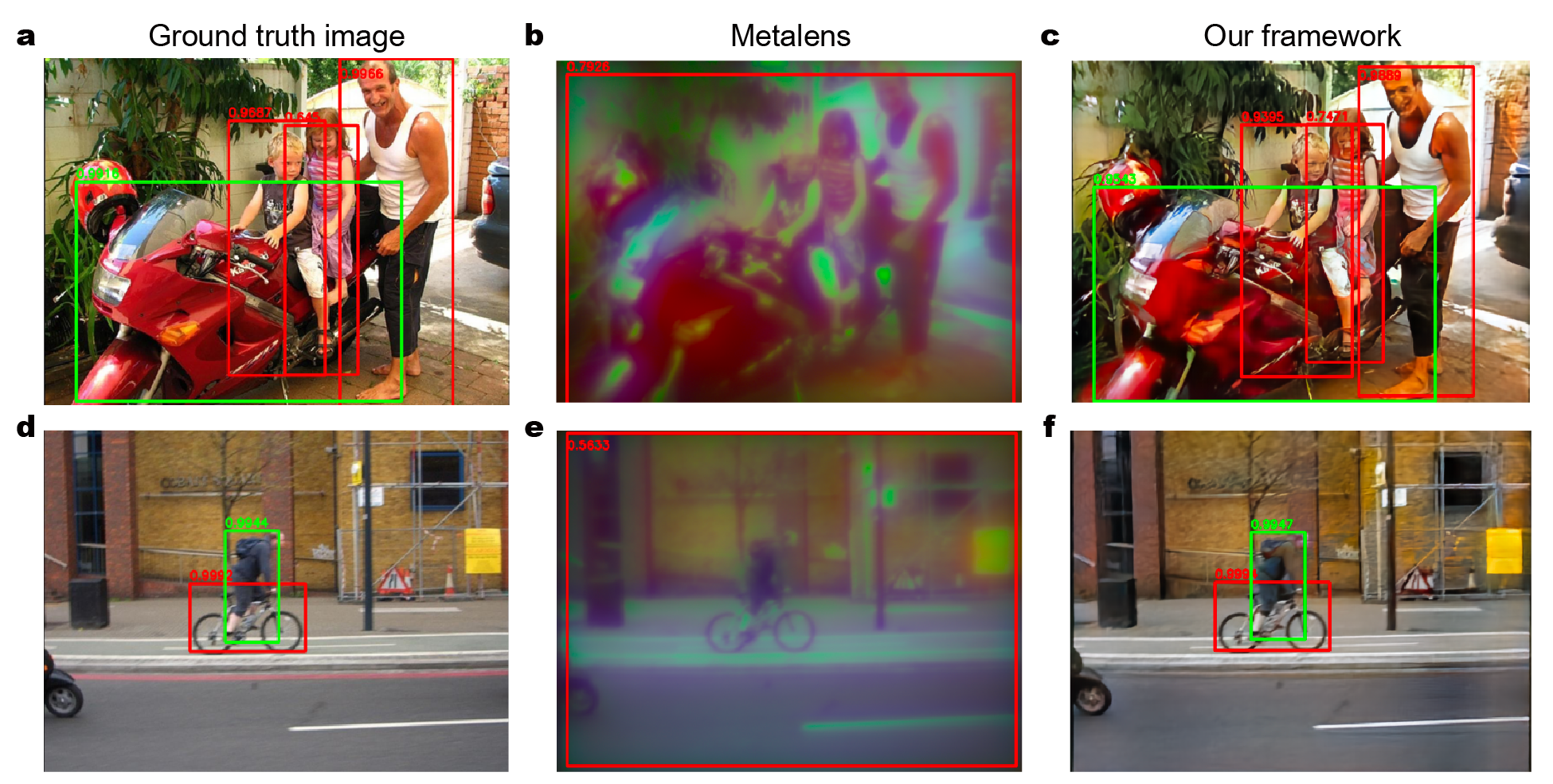}
    \caption{Object detection results using a pre-trained SSD model on the original images (a, d), the metalens images (b, e), and the images restored by our framework (c, f). The pre-trained SSD model could not detect any objects in the metalens images accurately; however, it successfully captured multiple classes and objects in images restored by our framework.} 
    \label{fig:fig7}
\end{figure}
We also assess the integrated system’s utility beyond image quality enhancement by transitioning to one of the domains of practical applications, object detection. To validate the performance of our framework in object detection on the restored images, we first obtained a test dataset consists of the ground truth, metalens, restored images using the entire PASCAL2007 dataset (n=4,952)~\cite{everingham2010pascal}. This dataset comprises 4,952 images with information of positions of objects and bounding box annotations for instances belonging to 20 different categories. Then, we employ Single Shot multibox Detector (SSD)~\cite{liu2016ssd} pre-trained for the PASCAL VOC2007 to detect bounding boxes of given images. Especially, we introduce mean average precision (AP) to evaluate object detection results. AP measures the model's accuracy in predicting the presence and correct localization of objects within an image. It provides a comprehensive assessment of the detection performance across varying thresholds of precision and recall, making it a standard benchmark in evaluating object detection algorithms. Specifically, AP is calculated in multiple scales using IoU thresholds ranging from 0.5 to 0.95. The terms ``$\mathrm{AP_{50}}$'' and ``$\mathrm{AP_{75}}$'' each represent the results of measurements with the IoU threshold set to 0.5 and 0.75, respectively.

Figure~\ref{fig:fig7} shows examples of object detection using SSD. The detector predicts the entire region (red box) as an object because it cannot identify the detailed features in the metalens images (Fig.~\ref{fig:fig6}(b)). On the other hand, the detector accurately predicts the bounding boxes at the desired objects in the restored images because compared to the original PASCAL VOC2007, the quality of the restored images is competitive (Fig.~\ref{fig:fig6}(a), (d) and (c), (f)). Our’s $\mathrm{AP}, \mathrm{AP_{50}}$ on the restored images are approximately 34\%p and 56\%p higher than those on the metalens images. They are approximately 86\% and 88\% of the $\mathrm{AP}, \mathrm{AP_{50}}$ on the ground truth (Table~S2). The improvement in AP scores for our framework-restored images compared to the original metalens images signifies the restoration's practical implications. Higher AP scores on our framework-restored images indicate that the model effectively recovers enough detail and structure from the aberrated metalens images to facilitate accurate object detection, closely approximating the performance on ground truth images. This enhancement is particularly crucial for applications in autonomous navigation, surveillance, and augmented reality, where precise object detection is paramount. 

\section{Conclusion}

In this study, we have demonstrated deep neural network (DNN)-based image restoration framework for large-area mass-produced metalenses. Our approach effectively mitigates the severe chromatic and angular aberrations inherent in large-area broadband metalenses, a challenge that has long impeded the widespread adoption of the metalenses. Also, assuming the uniform quality of mass-produced metalenses, the optimized restoration model can be applied to other metalenses manufactured at the same process. By employing an adversarial learning scheme in the Fourier space coupled with positional embedding, we have transcended traditional limitations, enabling the restoration of high-spatial-frequency information and facilitating aberration-free, full-color imaging through mass-produced metalenses. The profound implications of our findings extend a commercially viable pathway toward the development of ultra-compact, efficient, and aberration-free imaging systems.

\section{Data Availability} 
The metalens and ground truth image dataset can be accessed from the GitHub repository at \url{https://github.com/yhy258/EIDL_DRMI}~\cite{gitseo2023} and Figshare repository at \url{https://doi.org/10.6084/m9.figshare.24634740.v1}~\cite{seo2023}. USAF images taken with the metalens, reconstructed with the NAFNet, and reconstructed with our framework are also available in the GitHub repository. The test set of PASCAL VOC2007 dataset can be found at \url{http://host.robots.ox.ac.uk/pascal/VOC/voc2007/}.

\section{Code Availability} 
The code in this study is available in the GitHub repository at \url{https://github.com/yhy258/EIDL_DRMI}~\cite{gitseo2023}. The pre-trained models are also available in the same repository. We used the publicly available SSD code (\url{https://github.com/amdegroot/ssd.pytorch}) for object detection using the PyTorch library.

\section{Acknowledgements}
This work was supported by the National Research Foundation of Korea(NRF) grant funded by the Korea government(MSIT)) (RS-2023-00240713), and Artificial Intelligence Graduate School Program (No. 2020-0-01373, Hanyang University) supervised by the IITP (Institute for Information and Communications Technology Planning and Evaluation), and under the artificial intelligence semiconductor support program to nurture the best talents (IITP-2023-RS-2023-00253914) grant funded by the Korea government(MSIT), and a grant of the Korea Health Technology R\&D Project through the Korea Health Industry Development Institute (KHIDI), funded by the Ministry of Health \& Welfare, Republic of Korea (grant number: HI19C075300), and the POSCO-POSTECH-RIST Convergence Research Center programme funded by POSCO, and the NRS grant (NRF-2022M3C1A3081312, NRF-2019R1A5A8080290) funded by the MSIT of the Korean government.

\section{Author contributions}
H.C. and J.R. supervised and designed the study. J.S. and J.J. developed the methodology for the idea and performed the experiments with the assistance of C.K. and J.H.. J.J. acquired the data with assistance from J.S. Joohoon Kim, J.S., and J.J. analyzed the metalens and image data with assistance from S.M., E.L., and Joonho Kang. J.R. and Joohoon Kim designed and fabricated the metalenses. H.C., J.R., J.S., J.J., Joonho Kang, and Joohoon Kim wrote the paper.

\section{Conflict of interest}
The authors declare no competing interests.

\printbibliography

\end{document}


\maketitle
\tableofcontents
\newpage

\section{Metalens fabrication}

The metalens developed in this work comprises a two-dimensional rectangular array of meta-atoms, each featuring a nano-slab and an arbitrary rotation angle, forming a Pancharatnam-Berry phase-based metalens. The dimensions of the nano-slabs, which are 70 nm wide, 380 nm long, and 900 nm high, were established following the deposition of a 23 nm thick titanium dioxide thin film on the imprinted resin structures. The design parameters, including the width, length, height of the slab, and thickness of the titanium dioxide thin film, were systematically optimized to maximize the focusing efficiency. The metalens focuses light with the polarization-dependent transmittance of the meta-atom, the Jones matrix of which can be expressed as: 
\begin{equation}
    \textbf{J} = 
    \begin{bmatrix}
        t_l & 0 \\ 0 & t_t
    \end{bmatrix}
\end{equation} where \(t_l\) and \(t_t\) are the complex transmission coefficients of the meta-atoms when the electric field polarization is aligned along the longitudinal or transverse directions of the slab, respectively. 

The Jones matrix for the rotation angle \(\theta\) of the nano-slab is  
\begin{equation}
    \textbf{T} = \textbf{R}(-\theta) \textbf{J} \textbf{R}(+\theta) = 
    \begin{bmatrix}
        \cos\theta & \sin\theta \\ -\sin\theta & \cos\theta    
    \end{bmatrix}
    \begin{bmatrix}
        t_l & 0 \\ 0 & t_t    
    \end{bmatrix}
    \begin{bmatrix}
        \cos\theta & -\sin\theta \\ \sin\theta & \cos\theta    
    \end{bmatrix}
\end{equation} where \(\textbf{R}\) is the rotation matrix of the rotation angle \(\theta\), and \(\textbf{T}\) is the transfer matrix of the meta-atom. For incident Left-handed Circularly Polarized (LCP) light, the transmitted light can be derived as a linear combination of LCP and Right-handed Circularly Polarized (RCP) light: 
\begin{equation}
    \textbf{T} \frac{1}{\sqrt{2}}
    \begin{bmatrix}
        1 \\ i    
    \end{bmatrix}
    = 
    \frac{t_l+t_t}{2}
    \begin{bmatrix}
        1 \\ i
    \end{bmatrix}
    +
    \frac{t_l-t_t}{2} e^{i2\theta}
    \begin{bmatrix}
        1 \\ -i
    \end{bmatrix}
\end{equation}
The phase retardation of the RCP light can be precisely controlled by \(\theta\). The ideal phase distribution ($\phi_{\text{Ideal}}$) at the exit plane of the metalens is defined as 
\begin{equation}
    \phi_{\text{Ideal}}(x,y) = -\frac{2\pi}{\lambda} \left(
        \sqrt{x^2+y^2+f^2} - f
    \right)
\end{equation} where \(\lambda\) is the target wavelength of 532 nm; \(x\) and \(y\) are the spatial coordinates on the metalens; and \(f\) is the focal length, which was set to 24.5 mm for a numerical aperture of 0.2. Therefore, the orientation angle of each meta-atom at a point \((x,y)\) is ($\phi_{\text{Ideal}}/2$). 

To mass-produce the designed metalens on a wafer scale, we sequentially applied high-speed electron-beam lithography, ArF immersion scanning, nanoimprint lithography, and atomic layer deposition. The high-speed electron-beam lithography process (JBX Series, JEOL) was instrumental in patterning the photomasks of the metalenses. At this stage, the photomask contained a mask pattern for a singular metalens, which was insufficient for mass production. To rectify this, we transferred the photomask pattern onto a \ang{;;12} Si wafer in an array format using an ArF immersion scanner (XT-1900Gi, ASML), thereby creating a master stamp with metalens arrays.

Following this, we employed nanoimprint lithography to replicate the fabricated \ang{;;12} master stamp at a significantly reduced cost. In the initial phase of nanoimprint lithography, we coated the prepared master stamp with hard-polydimethylsiloxane (h-PDMS) to achieve a high-resolution replication of the metalens, and subsequently coated it with PDMS to create a replica mold. After baking the replica mold at 80°C for two hours, we separated the cured replica mold from the master stamp.

We then applied a conventional imprint resin (MINS-311RM) onto the replica mold and covered it with an \ang{;;4} glass wafer. Following the curing of the imprint resin under a pressure of 2 bar and UV-light irradiation, we formed a metalens pattern on the glass wafer by detaching the replica mold. To enhance the effective refractive index and thereby increase the efficiency further, we thinly coated the imprinted metalens with a high-index titanium dioxide film using atomic layer deposition. 

\section{Metalens imaging system and data acquisition setup}

\begin{figure}[H]
    \centering
    \includegraphics[width=1.0\linewidth]{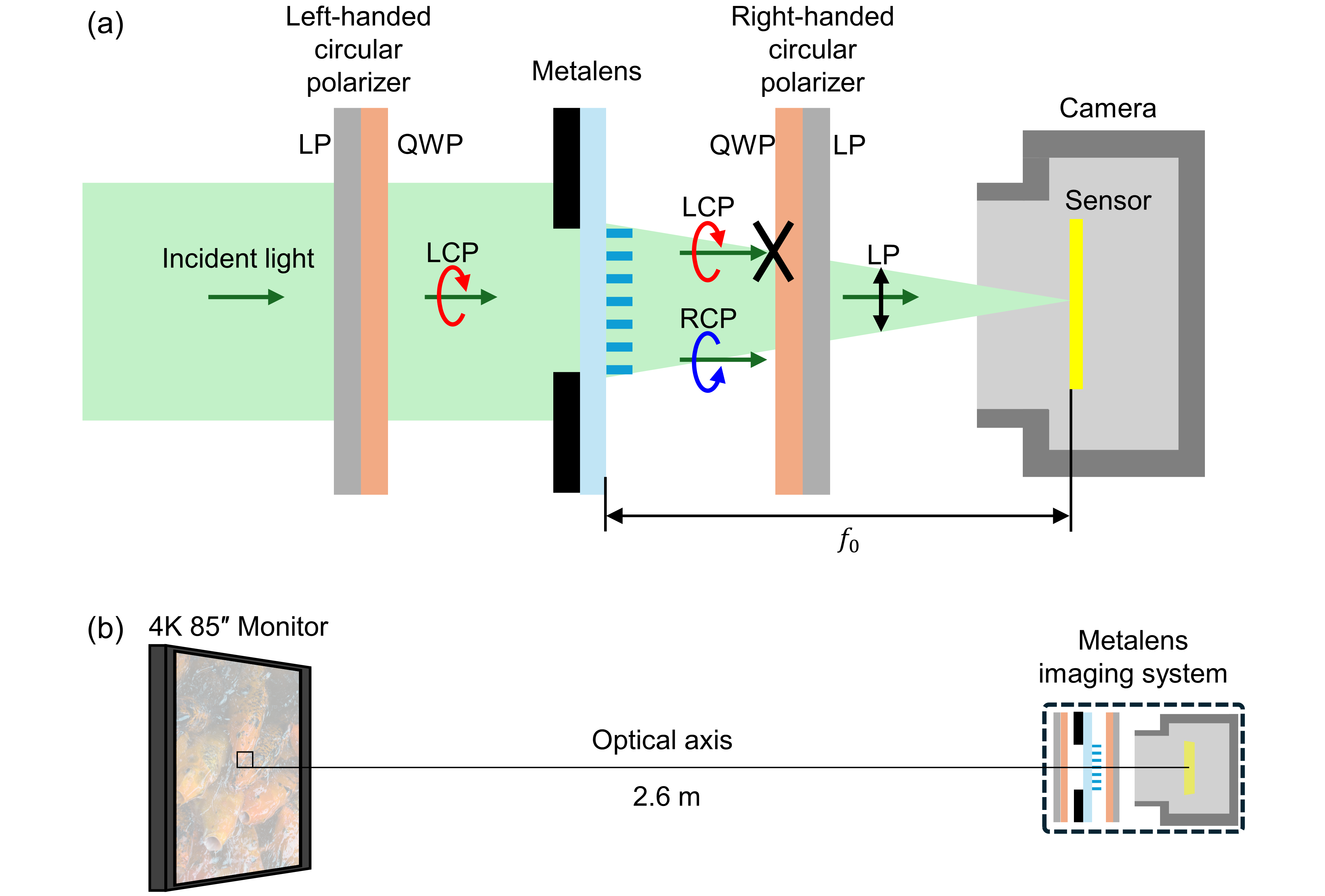}
    \caption{(a) Schematics of the metalens imaging system. The metalens imaging system consists of the metalens, the camera and left- and right-handed circular polarizers. (b) Schematics of data acquisition setup.}
    \label{fig:sup_fig1}
\end{figure}

The metalens imaging system incorporates our mass-produced metalens, a commercial camera (Basler acA5472-17uc) with an exposed imaging sensor and two left- and right-handed circular polarizers (Edmund Optics CP42HE and CP42HER). In addition, the polarizers can be expressed as stacks of a Linear Polarizer (LP) and a Quarter Wave Plate (QWP) because they are fabricated by laminating an XP42 LP sheet to a WP140HE QWP sheet. The roles of each optical component are shown in Fig.~S\ref{fig:sup_fig1}(a). The left-handed circular polarizer only transmits the LCP light to the metalens. As described in the metalens fabrication section, the RCP light transmitted from the metalens has the wavefront propagating to the focal point, while the LCP light has the same wavefront as the incident light. The right-handed circular polarizer transmits only RCP light and blocks LCP light to remove unfocused lights. Finally, the linearly polarized light transmitted from the right-handed circular polarizer focuses on the image sensor. 

The planes of the circular polarizers and image sensor are aligned perpendicular with the optical axis of the metalens, which point towards the center of the image sensor. Finally, the distance between the metalens and the image sensor f0 is set to 24.5 mm to focus the 532 nm green light. Figure~S\ref{fig:sup_fig1}(b) shows the data acquisition setup. The metalens imaging system is distant 2.6 m from the 4K \ang{;;85} monitor (Samsung KU85UA7050F) with the optical axis perpendicularly pointing toward the center of the monitor. The raw metalens images were obtained by capturing the ground truth images displayed on the monitor. 

\section{PSF measurement setup and MTF calculation}

\begin{figure}[H]
    \centering
    \includegraphics[width=1.0\linewidth]{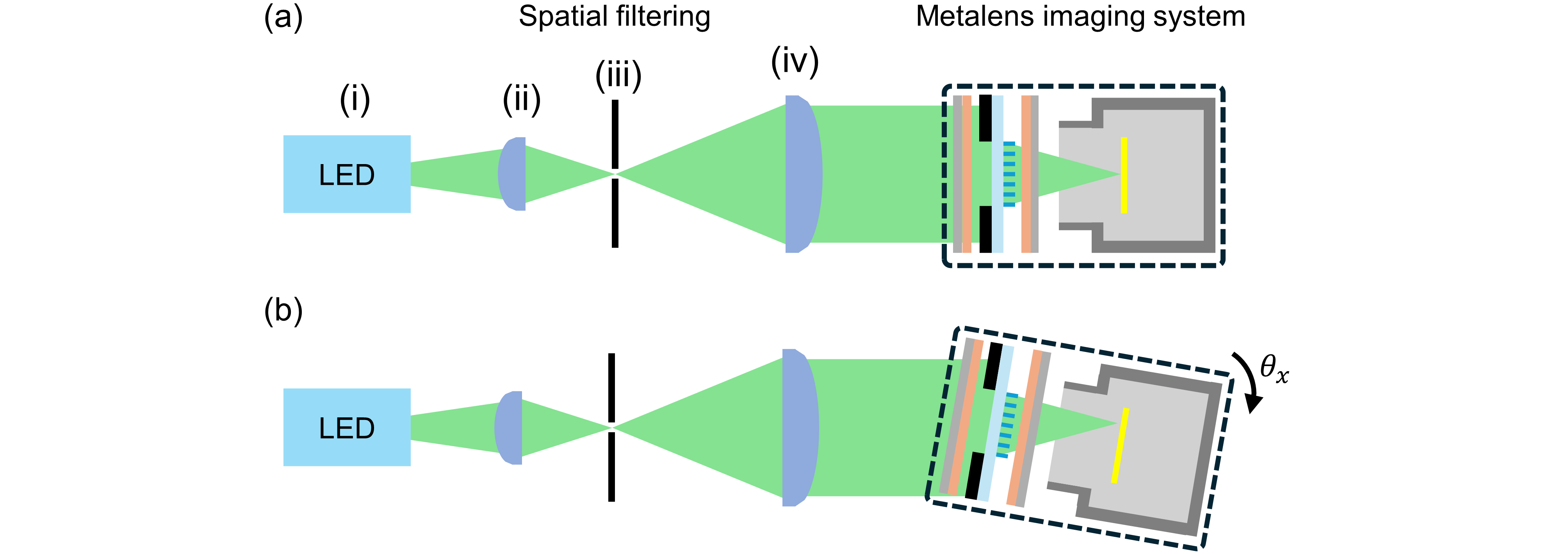}
    \caption{(a) Optical setup for PSF measurement with zero viewing angle. (b) PSF measurement setup with horizontal viewing angle $\theta_x$.}
    \label{fig:sup_fig2}
\end{figure}

The PSF was measured by capturing images of the collimated lights using the metalens imaging system. As described in Fig.~S\ref{fig:sup_fig2}(a), the red, green, or blue light from the light-emitting-diode with the peak intensities at 640, 525, 457 nm (i) is spatially filtered and collimated by the \ang{;;0.5} aspherical lens (ii, Thorlabs AC127-019-A), 20 um pinhole (iii, Thorlabs P20CB), and \ang{;;1} spherical lens (iv, LA1461-A). The PSFs with zero viewing angle were measured when the optical axes of the spatial filter and the metalens imaging system were matched. In addition, the PSFs with non-zero viewing angles were measured by tilting the metalens imaging system in the horizontal direction as shown in Fig.~S\ref{fig:sup_fig2}(b). 

The MTFs are subsequently calculated using measured PSFs as~\cite{goodman2005introduction},
\begin{equation} 
    \text{MTF} \equiv \left| \frac{\int\int I(x,y)\exp \left[-2\pi i(f_x x + f_y y)\right] dxdy}{\int\int I(x,y) dxdy} \right| 
    \label{eq:MTF} 
\end{equation}
where $x$ and $y$ are the horizontal and vertical positions on the image sensor; $I(x,y)$ is the PSF; and $f_x$ and $f_y$ are the spatial-frequencies along the $x$ and $y$ axes, respectively. Defining the viewing angle as the angle between the optical axis of the metalens and the line towards the point light source from the center of the metalens, Fig.~\ref{fig:sup_fig2}(d) describes the MTFs depending on $f_x$ with the zero $f_y$.

\section{Statistical train-test inconsistency}

\begin{figure}[H]
    \centering
    \includegraphics[width=1.0\linewidth]{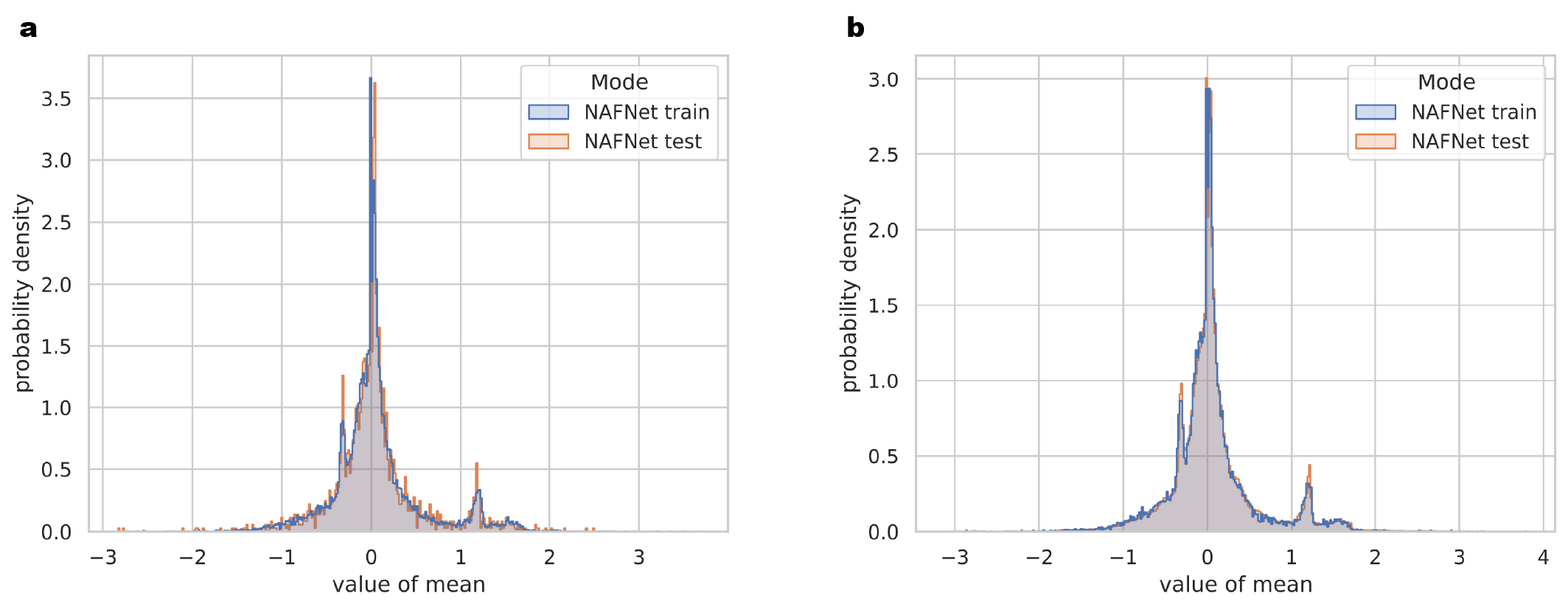}
    \caption[Statistical train-test inconsistency]%
    {\textbf{Statistical train-test inconsistency}. (a-b) Probability density functions of global average pooled features of training(red) and testing(blue) (a) without and (b) with TLC.
    }
    \label{fig:sup_fig3}
\end{figure}

In natural image restoration studies based on deep neural networks (DNNs)~\cite{cui2023selective, Zamir2022MIRNetv2, Chen_2021_CVPR}, the channel attention module~\cite{zhang2018image} leads to the exceptional performance of the restoration. Global average pooling (GAP) shrinks the given features in spatial dimensions to generate channel-wise statistics. However, GAP significantly degraded the train-test consistency because we used randomly cropped image patches as training data and full-resolution images (non-cropped) as test data, as mentioned in the main text. Figure~S\ref{fig:sup_fig3} shows this statistical inconsistency at the training and test phases of NAFNet (baseline model). Specifically, Figures~S\ref{fig:sup_fig3}(a) and (b) show the probability density functions of the outputs of the GAP of the first channel attention module at the second encoder of the model for the metalens images. The test-time local converter (TLC)\cite{chu2022improving} can significantly contribute to overcoming these issues. After applying TLC, the statistics in the test phase are similar to those in the training phase (Fig.~S\ref{fig:sup_fig3}(b)). Furthermore, TLC ensures this statistical consistency and significantly improves the image quality (Table~S\ref{tab:sup_tab2}).

\section{Pattern artifact resulting from adversarial learning scheme in RGB space}

\begin{figure}[H]
    \centering
    \includegraphics[width=1.0\linewidth]{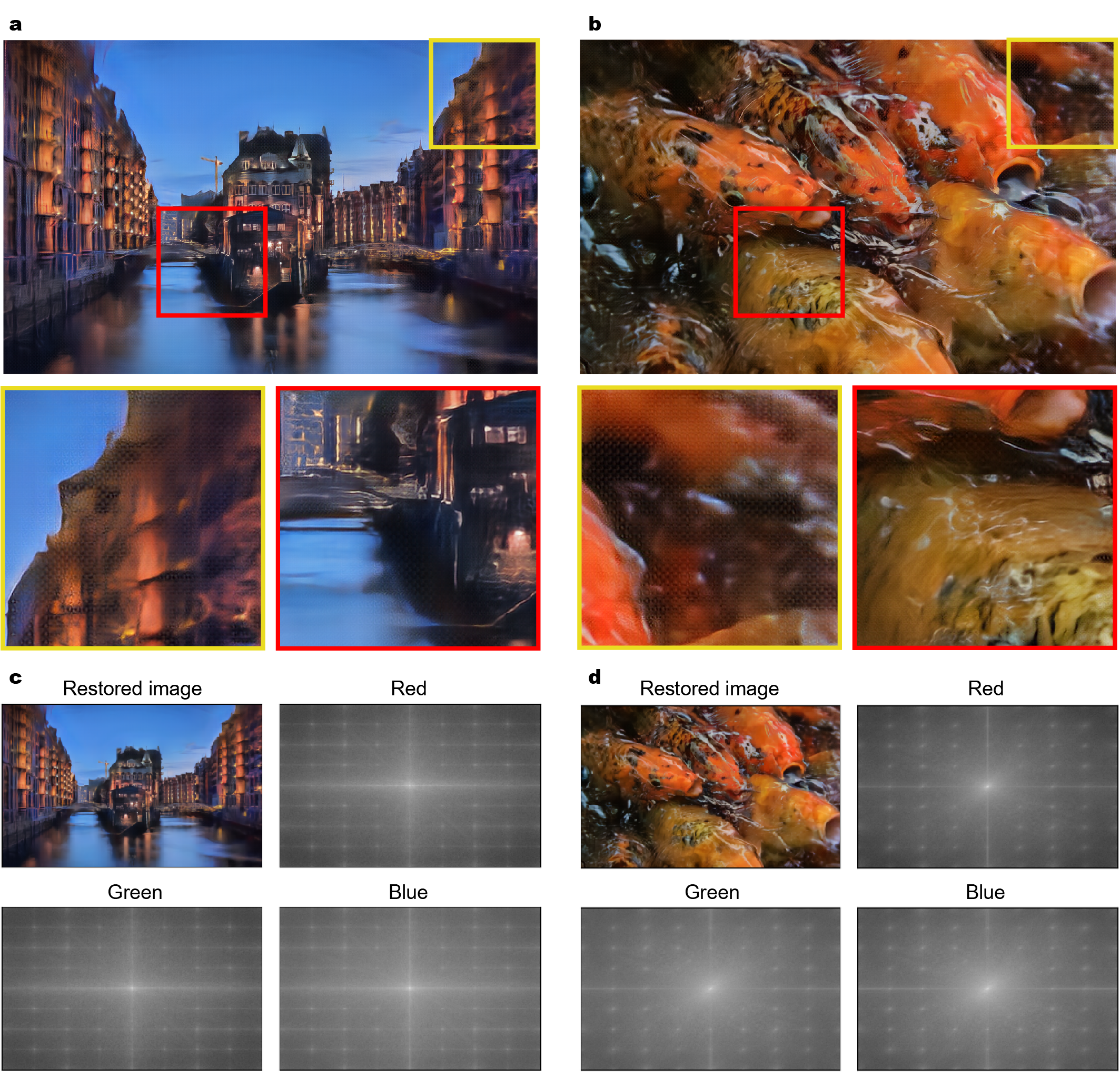}
    \caption{Pattern artifacts resulting from the adversarial learning scheme in RGB space. (a-b) Results of image restoration on the test set using adversarial learning regularization in RGB space. (c-d) Visualization of (a-b) in the frequency domain.
    }
    \label{fig:sup_fig4}
\end{figure}

As mentioned in the main text, the implementation of an adversarial learning scheme in the RGB space for regularization results in restored images that exhibit pattern artifacts in the RGB space across the entire area (Fig.~S\ref{fig:sup_fig4}(a) and (b)) and the frequency domain (Fig.~S\ref{fig:sup_fig4}(c) and (d)). In particular, applying adversarial learning in the RGB space significantly improved LPIPS but drastically reduced PSNR and SSIM (PSNR : 21.48 versus 21.69; SSIM : 0.64 versus 0.68; LPIPS : 0.39 versus 0.44).

\section{Details of Performance Metrics }

\subsection{PSNR and SSIM }
We use PSNR and SSIM, representative image quality metrics, to evaluate the images reconstructed by the proposed model. The PSNR is : 

\begin{equation} 
\mathrm{PSNR}(x, \hat x) = 10 \log_{10}({R^2\above 1pt \mathrm{MSE}(x, \hat x)}). 
\end{equation}Since MSE is a metric indicating the difference of pixel values, higher PSNR values indicate a better restoration performance. \(R\) is the maximum signal value of the ground truth image. However, PSNR is only an approximation to the human visual perception of the reconstruction quality because MSE is the pixel-level intensity difference between the reconstructed and ground truth images. Therefore, we use SSIM to accurately evaluate the visually perceived quality of the restored images. SSIM utilizes three elements in its evaluation: luminance, contrast, and structure between the reconstructed and ground truth images. The SSIM is : 

\begin{equation} 
\mathrm{SSIM}(\hat x , x) = {(2\mu_{\hat x}\mu_x + c_1)(\sigma_{\hat x x}+c_2) \over(\mu_{\hat x}^2 + \mu_x^2 + c_1)(\sigma_{\hat x}^2 + \sigma_x^2 + c_2)} 
\end{equation} In this equation, \(\mu_x\), \(\sigma^2_x\) and \(\sigma^2_{xy}\) indicate mean of \(x\), variance of \(x\) and covariance of \(\hat x\), \(x \), respectively. 

\subsection{Assessment in Spatial Frequency Domain}
We also introduce metrics to evaluate the restoration of lost spatial frequency information due to intense degradation in the Fourier domain. Spatial frequency information is entirely characterized by the magnitude and phase angle. Thus, we employ magnitudes’ MAE and cosine similarity as metrics in the spatial frequency domain. First, we transform the given ground truth and restored images from RGB space into Fourier space. Then, we derive the MAE and cosine similarity between the magnitudes of the transformed data. The MAE and cosine similarity are:  

\begin{equation} 
\mathrm{MAE}_{\mathcal{F}}(\hat x,x) = {1 \over N} \sum^N_{n=1}| |\mathcal{F}(\hat x)| - |\mathcal{F}(x)|| 
\end{equation} 

\begin{equation} 
CS_\mathcal{F}(\hat x, x) = {1\over N}\sum^N_{n=1}{\mathcal{F}(\hat x_n) \cdot \mathcal{F}(x_n)\over ||\mathcal{F}(\hat x_n)||_2||\mathcal{F}(x_n)||_2} 
\end{equation}In this equation, \(\mathcal{F}\) is Fourier transformation.

\newpage

\begin{table}[H]
    \centering
    \begin{tabular}{lllll}
    \toprule
                       & TLC & PSNR   & SSIM  & LPIPS \\
    \midrule
    \multirow{2}{*}{HINet}  &     & 19.378 & 0.645 & 0.493 \\
    \cmidrule(rl){3-5}
                            & \checkmark  & 21.364 & 0.677 & 0.456 \\
    \midrule
    \multirow{2}{*}{NAFNet} &     & 18.447 & 0.588 & 0.519 \\
    \cmidrule(rl){3-5}
                            & \checkmark  & 21.689 & 0.676 & 0.440 \\
    \midrule
    \multirow{2}{*}{Our framework}   &     & 20.868 & 0.678 & 0.448 \\
    \cmidrule(rl){3-5}
                            & \checkmark  & 22.095 & 0.692 & 0.432 \\
    \bottomrule
\end{tabular}
\caption{\label{tab:sup_tab1} Comparison of quantitative results from various models without and with TLC.}
\end{table}

\begin{table}[H]
    \centering
    \begin{tabular}{llll}
    \toprule
                        & \(\mathrm{AP}\)   &  \(\mathrm{AP_{50}}\)  &  \(\mathrm{AP_{75}}\) \\
    \midrule
    Ground truth    & 0.451 & 0.730 & 0.482 \\
    Metalens image  & 0.051 & 0.087 & 0.054 \\
    Our framework    & 0.386 & 0.646 & 0.397 \\
    \bottomrule
\end{tabular}
\caption{\label{tab:sup_tab2}  Comparison of quantitative results of object detection on the PASCAL VOC2007 between ground truth images, metalens images, and restored images.}
\end{table}

\printbibliography